\documentclass[aps,prl,reprint,superscriptaddress,longbibliography,showpacs,showkeys]{revtex4-1}

\usepackage{amsmath,bm,bbm,amssymb}
\usepackage[squaren]{SIunits}
\usepackage[usenames,dvipsnames]{color}
\usepackage[pdftex]{graphicx}
\usepackage{ulem}
\usepackage{hyperref}
\usepackage[english]{babel}

\newcommand{\ave}[1]{\ensuremath{\langle#1\rangle}}

\newcommand{\cov}{\mathrm{cov}}

\newcommand{\Tr}{\mathrm{Tr}}

\newcommand{\Fx}{F_x}
\newcommand{\Fy}{F_y}
\newcommand{\Fz}{F_z}

\newcommand{\FPlane}{F_{||}}

\newcommand{\Sx}{S_x}
\newcommand{\Sy}{S_y}
\newcommand{\Sz}{S_z}
\newcommand{\Sk}{S_k}

\newcommand{\NA}{N_{\rm A}}
\newcommand{\NL}{N_{\rm L}}

\newcommand{\nl}{n_{\rm l}}

\newcommand{\nlh}{n_{\rm l}^{(H)}}

\newcommand{\rb}{$^{87}\mathrm{Rb}\ $}

\newcommand{\F}{\mathbf{F}}
\newcommand{\f}{\mathbf{f}}

\newcommand{\tpoint}{t_k}
\newcommand{\tdelta}{\Delta t}
\newcommand{\tmeas}{t_{\rm e}}
\newcommand{\trel}{t_{\rm r}}

\newcommand{\xiPlane}{\xi_{||}^2}
\newcommand{\xiPlaneM}{\xi_{\rm m}^2}
\newcommand{\xiPlaneE}{\xi_{\rm e}^2}
\newcommand{\xiY}{\xi_{y}^2}
\newcommand{\xiZ}{\xi_{z}^2}

\newcommand{\offset}{\varphi_0}

\newcommand{\etaSc}{\eta_{\rm sc}}
\newcommand{\etaDec}{\eta_{\rm dec}}

\newcommand{\omegaL}{\omega_{\rm L}}

\begin{document}

\title{Entanglement-enhanced phase estimation without prior phase information}

\newcommand{\ICFOAddress}{ICFO-Institut de Ciencies Fotoniques, The Barcelona Institute of Science and Technology, 08860 Castelldefels (Barcelona), Spain}
\newcommand{\ICREAAddress}{ICREA -- Instituci\'{o} Catalana de Re{c}erca i Estudis Avan\c{c}ats, 08015 Barcelona, Spain}
\newcommand{\BilbaoAddress}{Department of Theoretical Physics, University of the Basque Country UPV/EHU, P.O. Box 644, E-48080 Bilbao, Spain}

\author{G.~Colangelo}
\email[]{giorgio.colangelo@icfo.es}
\affiliation{\ICFOAddress}

\author{F.~Martin Ciurana}
\affiliation{\ICFOAddress}

\author{G.~Puentes}
\affiliation{Departamento de F\'{i}sica, Facultad de Ciencias Exactas y Naturales, Pabell\'{o}n 1, Ciudad Universitaria, 1428 Buenos Aires, Argentina}

\author{M.~W.~Mitchell}
\email[]{morgan.mitchell@icfo.es}
\affiliation{\ICFOAddress}
\affiliation{\ICREAAddress}

\author{R.~J. Sewell}
\email[]{robert.sewell@icfo.eu}
\affiliation{\ICFOAddress}

\date{\today}

\begin{abstract}
We study the generation of planar quantum squeezed (PQS) states  by quantum non-demolition (QND) measurement of a cold ensemble of $^{87}$Rb atoms.  Precise calibration of the QND measurement allows us to infer the conditional covariance matrix describing the $\Fy$ and $\Fz$ components of the PQS, revealing the dual squeezing characteristic of PQS.  PQS states have been proposed for single-shot phase estimation without prior knowledge of the likely values of the phase. We show that for an \textit{arbitrary} phase, the generated PQS gives a metrological advantage of at least $\unit{3.1}{dB}$ relative to classical states. The PQS also beats traditional squeezed states generated with the same QND resources, except for a narrow range of phase values.  Using spin squeezing inequalities, we show that spin-spin entanglement is responsible for the metrological advantage.  
\end{abstract}


\maketitle

Estimation of interferometric phases is at the heart of precision sensing, and is ultimately limited by quantum statistical effects~\cite{WisemanBook2010}.  
Entangled states can improve sensitivity beyond the ``classical limits'' that restrict sensing with independent particles, and a diversity of entangled states have been demonstrated for this task, including photonic squeezed states \cite{SlusherPRL1985,WuPRL1986} and spin-squeezed states~\cite{MeyerPRL2001}.
These give improved sensitivity for a narrow range of phases, but worsened sensitivity for most phases.  
Optical ``NOON'' states~\cite{MitchellN2004} give improved sensitivity over the whole phase range, but introduce additional phase ambiguity that increases with the size, and thus sensitivity advantage, of the NOON state. 
Recent proposals~\cite{Toth2009,HePRA2011,HeNJP2012} suggest using \textit{planar quantum squeezed} (PQS) states to obtain an entanglement-derived advantage for all phase angles, with no additional phase ambiguity. 
A natural application is in high-bandwidth atomic sensing~\cite{ShahPRL2010,VasilakisPRL2011,SewellPRL2012}, in which the precession angle may not be predictable in advance.
PQS states may also be valuable for \textit{ab initio} phase estimation using feedback~\cite{XiangNPhot2010,YonezawaScience2012,BerniNatPhoton2015}.

Discussion of such states under the name ``intelligent spin states''~\cite{AragoneJPA1974} predates modern squeezing terminology, and analogous states have been studied with optical polarization~\cite{KorolkovaPRA2002,SchnabelPRA2003,Predojevic2008}.   
Generation of PQS states in material systems has been proposed using two-well Bose-Einstein condensates with tunable and attractive interactions~\cite{HePRA2011,HeNJP2012}, and  using quantum non-demolition (QND) measurements~\cite{PuentesNJP2013}.  
Here we take the latter approach, using Faraday rotation QND measurements~\cite{MitchellNJP2012,SewellNP2013} applied to an ensemble of cold atomic spins with $f=1$. As the ensemble spin precesses about the $x$ axis in an external magnetic field~\cite{BehboodPRL2013,Behbood2013APL,BehboodPRL2014}, we measure the $y$ and $z$ spin components to generate measurement-induced squeezing in these two components, creating a PQS state.  
The resulting state has enhanced sensitivity to precession angle, i.e., to Zeeman-shift induced phase. 
The demonstrated PQS state beats the best possible classical state at any precession angle, and beats traditional spin-squeezed states when averaged over the possible angles.  
Spin-squeezing inequalities~\cite{HePRA2011,HeNJP2012,VitaglianoPRL2011} detect spin entanglement in the PQS state, showing the sensing advantage is due to spin entanglement \cite{BeduiniPRL2015}.

A spin $\mathbf{F}$ obeys the Robertson uncertainty relation 
\begin{equation}
\label{eq:Robertson}
\Delta \Fy \Delta \Fz \ge \frac{1}{2} | \langle [\Fy, \Fz] \rangle | = \frac{1}{2} | \ave{\Fx} |.
\end{equation}
Unlike the canonical Heisenberg uncertainly relation, the rhs of Eq.~(\ref{eq:Robertson}) may vanish, e.g. for $\ave{\Fx}=0$, with the consequence that two spin components, e.g. $\Fy$ and $\Fz$, may be {\it simultaneously} squeezed, with the uncertainty absorbed by the third component, $\Fx$.
We refer to a state fulfilling this condition as a PQS state.

Following the approach of He {\it et al.}~\cite{HePRA2011,HeNJP2012}, we adopt an operational definition planar squeezing.
We take  $\Delta^2 \Fy = \Delta^2 \Fz = \FPlane/2$ as the standard quantum limit, where $\FPlane \equiv \sqrt{\Fy^2 + \Fz^2}$, so that $\FPlane$ is the magnitude of the in-plane spin components.  
We define the planar variance $\Delta^2 \FPlane \equiv  \Delta^2 \Fy + \Delta^2 \Fz$, with standard quantum limit $\Delta^2 \FPlane = \FPlane$, and  the planar squeezing parameter
\begin{equation}
\label{eq:PQSParameter}
\xiPlane \equiv \frac{\Delta^2 F_{||}}{\FPlane}.
\end{equation}
A PQS state has $\xiPlane < 1$, and has individual component variances below the standard quantum limit, i.e.,  {$ \xiY <  1$, and $ \xiZ < 1$ }, where $\xi_i^2 \equiv {2\Delta^2 F_{i}}/{\FPlane}$, so that $\xiPlane = (\xiY + \xiZ)/2$. 

Entanglement is detected using the witness $\xiPlaneE\equiv\Delta^2 F_{||}/\ave{\tilde{\NA}}$, derived in Ref.~\cite{HePRA2011}; for $f=1$ atoms, entanglement is detected if $\xiPlaneE<7/16$.
Here $\tilde\NA\equiv(\etaSc+p(1-\etaSc))\NA$ is the number of atoms remaining in the $f=1$ state after probing, $\etaSc$ accounts for off-resonant scattering of atoms, and $p$ is the fraction of scattered atoms that return to $f=1$~\cite{ColangeloNJP2013}.
We also define a metrological squeezing parameter $\xiPlaneM \equiv F \Delta^2 F_{||}/\FPlane^2$, where $F\equiv\ave{\NA}$ is the input spin coherence, similar to the Wineland criterion~\cite{WinelandPRA1992,WinelandPRA1994}, in that it compares noise to the magnitude of the coherence $\FPlane$.
A PQS with $\xiPlaneM<1$ gives enhanced metrological sensitivity to arbitrary phase shifts.

\begin{figure}[t]
\centering
\includegraphics[width=\columnwidth]{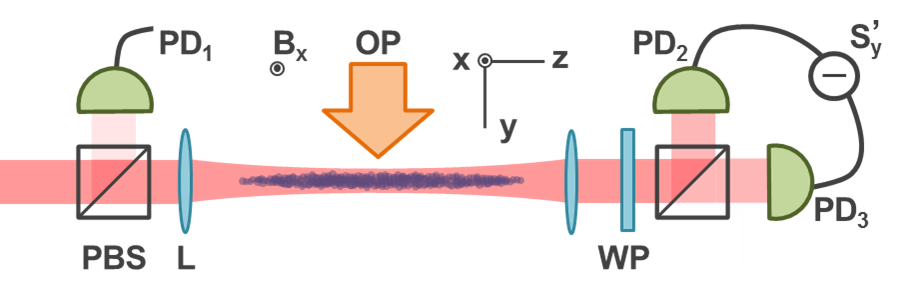}
\caption{
Experimental setup.
A cloud of laser-cooled $^{87}$Rb atoms is held in a singe-beam optical dipole trap.
The atoms precess in the $y$--$z$ plane due to an external magnetic  field $B_x$.
Off-resonant optical probe pulses experience Faraday rotation as they pass through the atoms by an angle $\varphi$ proportional to the collective on axis spin component $\Fz$. 
Rotation of the optical polarization from $\Sx$ into $\Sy'$ is detected by a balanced polarimeter that consists in a wave plate (WP), a polarizing beam splitter (PBS), and photodiodes PD$_{2}$ and PD$_3$.
The input $\Sx$ polarization is recorded with a reference photodetector (PD$_1$).
\label{fig:exp}}
\end{figure}

A PQS state may be used to measure arbitrary phase angles with quantum-enhanced precision.
For example, we consider an ensemble of atomic spins precessing in the $y$--$z$ plane in an external magnetic field $B_x$.
The spin projection onto the $z$-axis is given by $\Fz(t)=\Fz \cos \phi -\Fy \sin \phi$, 
where $\Fy$ and $\Fz$ are evaluated at $t=0$ and the phase $\phi=\omegaL t$ is proportional to the magnetic field.
The uncertainty in estimating $\phi$ of the atomic precession is
\begin{align}
	\Delta^2 \phi &= \frac{\Delta^2 \Fz(\phi)}{|d \ave{\Fz(\phi)}/d \phi|^2} = \frac{\Delta^2 \Fz(\phi)}{(\ave{\Fy}\cos\phi+\ave{\Fz}\sin\phi)^2}
\label{eq:phase}
\end{align}
where $\Delta^2 \Fz(\phi)\equiv \Delta^2 \Fy \sin^2\phi +\Delta^2 \Fz\cos^2\phi + \mathrm{cov}(\Fy,\Fz)\sin2\phi$, and $ \cov(A,B) \equiv  \frac{1}{2} \ave{A B + BA} - \ave{A}\ave{B}$ is the covariance.
The standard quantum limit is $\Delta^2\phi_{\rm SQL}=1/2 \FPlane$.
We note that PQS states reduce the planar variance for arbitrary angles on a finite interval, except where the denominator in Eq.~(\ref{eq:phase}) is equal to zero.
In contrast, squeezing a single spin component is only beneficial to refine the estimate of a phase over a limited range of angles, and requires prior knowledge of the phase, or adaptive procedures to determine the phase during the measurement~\cite{HeNJP2012}.

We work with an ensemble of up to $1.75\times 10^6$ laser-cooled \rb atoms held in a single beam optical dipole trap~\cite{KubasikPRA2009,KoschorreckPRL2010a,KoschorreckPRL2010b}, as illustrated in Fig.~\ref{fig:exp}.
The atoms are initially polarized via high efficiency ($\sim98\%$) stroboscopic optical pumping, in the presence of a small magnetic field applied along the $x$-axis, such that $\ave{\Fy}\simeq\ave{\NA}$.
$\NA$ is subject to Poissonian fluctuations because accumulation of independent atoms into the ensemble is a stochastic process limited by Poisson statistics $\Delta^2\NA = \ave{\NA}$.
We refer to this kind of state as a {\it Poissonian coherent spin state} (PCSS), with variances $\Delta^2\Fx = \Delta^2\Fz = \ave{\NA}/2$ and $\Delta^2\Fy = \ave{\NA}$.
Generating sub-Poissonian atom number statistics, either via strong interaction among the atoms during accumulation~\cite{SchlosserN2001,SortaisPRA2012,ChuuPRL2005,ItahPRL2010,SannerPRL2010,WhitlockPRL2010,HofmannPRL2013}, or precise non-destructive measurement~\cite{StocktonThesis2007,TakanoPRL2009,AppelPNAS2009,SchleierSmithPRL2010,HumePRL2013,BeguinPRL2014,BohnetNPhot2014,GajdaczPRL2016,HostenN2016,ZhangPRL2012,StroescuPRA2015}, remains a significant experimental challenge.

\begin{figure}[t]
\centering
\includegraphics[width=\columnwidth]{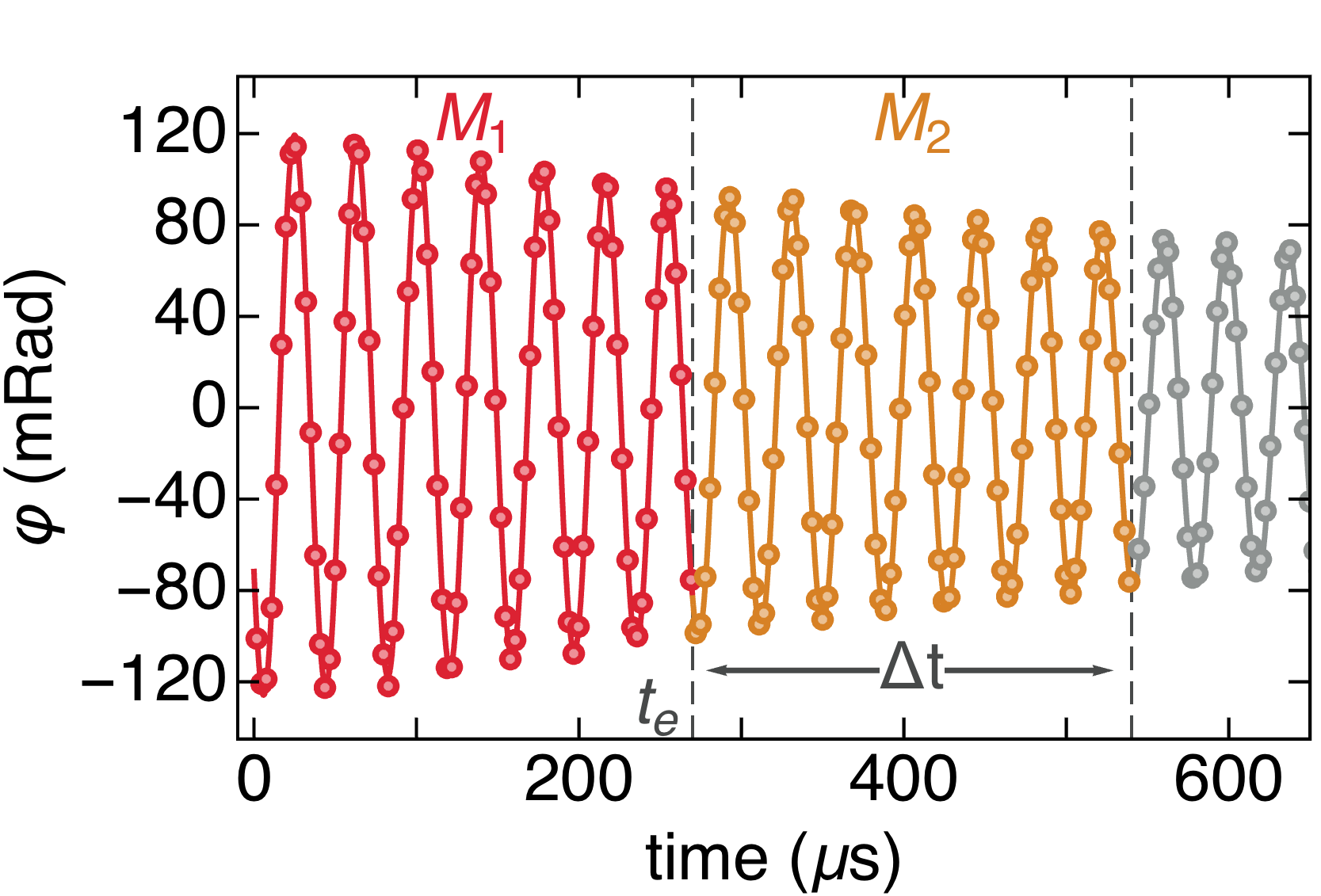}
\caption{
Rotation angle $\varphi$ in the $y-z$ plane of a $\Fy$-polarized state precessing under a magnetic field oriented in the $x$ direction. 
We use the measurement record to predict the $\Fz$ and $\Fy$ components at a time $t=\tmeas$ using two sequential measurements $M_1$ and $M_2$.
\label{fig:FID}
}
\end{figure}

We probe the atomic spins via off-resonant paramagnetic Faraday-rotation.
The effective atom-light interaction is given by the hamiltonian
\begin{equation}
H_{\rm eff} = g \Sz \Fz
\label{qnd}
\end{equation}
Here, the atoms are described by the collective spin operators $\F \equiv \sum_{i} \f^{(i)}$, with $\f^{(i)}$ the spin orientation of individual atoms.
The optical polarization of the probe pulses is described by the Stokes operators $\Sk = \frac{1}{2}(a_L^\dagger, a_R^\dagger) \sigma_k (a_L, a_R)^T$, with Pauli matrices $\sigma_k$.
The coupling constant $g$ depends on the detuning from the resonance of the probe beam, the atomic structure, the geometry of the atomic ensemble and probe beam~\cite{KubasikPRA2009,KoschorreckPRL2010a,Deutsch2010OC,KuzmichPRL2000,AppelPNAS2009}.

\begin{figure*}[t]
\includegraphics[width=\textwidth]{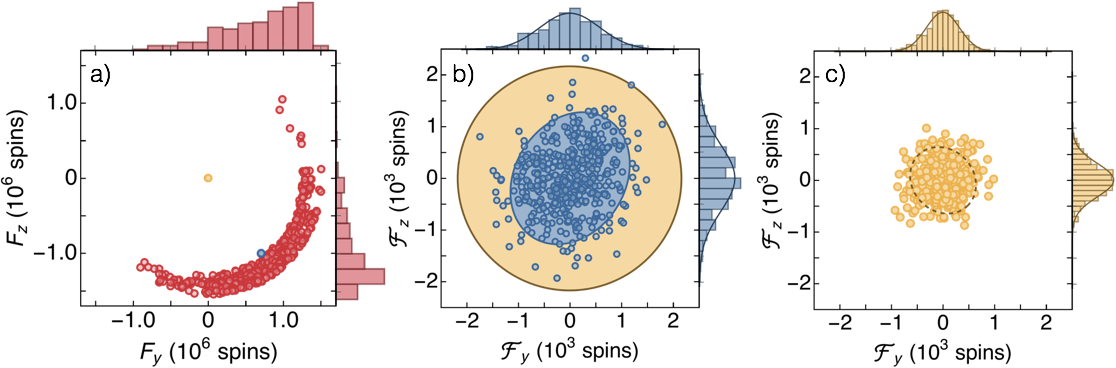}
\caption{
a) Spin state $\F_1$ (red dots) estimated at time $\tmeas$ for an input state with $\ave{\NA}=1.88\times10^6$ atoms from the 450 repetitions of the experiment. 
For comparison, we illustrate on the same scale the $\bm{\mathcal{F}}$, the best linear prediction of $\F_2$ given $\F_1$, around the mean vector $\ave{\F_1}$ (blue dot), and the corresponding measurement made without atoms in the trap, used to quantify the read-out noise (yellow dot).
b) Error in the best linear predictor, $\bm{\mathcal{F}}$ (blue dots).
The blue ellipse shows the measured $2\sigma$ radii of the distribution.
The yellow ellipse shows the standard quantum limit $\Delta^2 \Fy = \Delta^2 \Fz = \FPlane/2$ with $2\sigma$ radii, where $\sigma^2=(\FPlane/2)^2+\Delta^2\offset$ and $\Delta^2\offset$ is the measured read-out noise.
c) Linear predictor ${\bf \mathcal{F}}$ from repetition of the experiment without atoms in the trap, allowing quantification of the measurement read-out noise.
The dashed ellipse shows the measured $2\sigma$ radii of the distribution.
\label{fig:PQS}
}
\end{figure*}

Equation~\eqref{qnd} describes a quantum non-demolition measurement of the collective atomic spin $\Fz$: an input $\Sx$-polarized optical pulse interacting with the atoms experiences a rotation by an angle $\varphi = g \Fz$.
The transformation produced by the interaction is $\Sy' = \Sy \cos \varphi + \Sx \sin \varphi $.
In our experiment we measure $\Sx$ at the input by picking off a fraction of the optical pulse and sending it to a reference detector, and $\Sy'$ using a fast home-built balanced polarimeter~\cite{MartinOL2016}. 
Both signals are recorded on a digital oscilloscope, from which we calculate $\hat{\varphi}=\arcsin \left(\Sy'/\Sx \right)$, the estimator for $\varphi$.
We correct for slow drifts in the polarimeter signal by subtracting a baseline from each pulse, estimated by repeating the measurement without atoms in the trap.

We probe the atoms using a train of $\tau=\unit{0.6}{\micro\second}$ duration pulses of linearly polarized light, with a detuning of \unit{700}{\mega Hz} to the red of the \rb D$_2$ line, sent through the atomic cloud at \unit{3}{\micro\second} intervals.
The probe pulses are $V$-polarized, with on average $\nl=2.74 \times 10^6$ photons.
Between the probe pulses, we send $H$-polarized compensation pulses with on average $\nlh= 1.49 \times  10^6$ photons through the atomic cloud to compensate for tensor light shifts~\cite{KoschorreckPRL2010b,SewellNP2013,ColangeloNJP2013}.
During the measurement, an external magnetic field $B_x$ coherently rotates the atoms in the $y$--$z$ plane at the larmor frequency $\omegaL$.
The time taken to complete a single-pulse measurement is small compared to the Larmor precession period, i.e. $\tau \ll T_L$.
Off-resonant scattering of probe photons during the measurement leads to decay of the atomic coherence at a rate $\eta=3\times10^{-10}$ per photon.

The measurable signal is described by the free induction decay model~\cite{Behbood2013APL}
\begin{equation}
\label{eq:FIDForm}
\varphi(t)=g \Big( \Fz(\tmeas) \cos \phi -\Fy(\tmeas) \sin \phi \Big) e^{-\trel/T_2} + \offset
\end{equation}
where $\trel \equiv t- \tmeas$ and the phase $\phi=\omegaL \trel$ is proportional to the magnetic field.
We record a set of measurements $\varphi(\tpoint)$, and detect the PQS state at time $\tmeas$.
A typical free induction decay signal is illustrated in Fig.~\ref{fig:FID}.
An independent measurement is used to calibrate $g$, while $\omegaL$, $T_2$, and $\offset$ are found by fitting  the measured $\varphi(\tpoint)$ over all the  data points.

The model described in Eq.~\eqref{eq:FIDForm} allows a simultaneous estimation of $\F_1 = (\Fy^{(1)}, \Fz^{(1)})$ at a time $t=\tmeas$ by fitting the the data using the measurements from an interval $\tdelta$ \textit{prior} to $\tmeas$ (labeled $M_1$ in Fig.~\ref{fig:FID}), producing a conditional PQS at time $\tmeas$.
We detect the PQS by comparing the first measurement outcome to a second estimate $\F_2 = (\Fy^{(2)}, \Fz^{(2)})$ using the measurements from an interval $\tdelta$ \textit{after} to $\tmeas$ (labeled $M_2$ in Fig.~\ref{fig:FID}).
The classical parameters $g$, $\omegaL$, $T_2$ and $\offset$ are fixed beforehand.
As a result, these are two linear, least-squares estimates of the vector $\F$ obtained from disjoint data sets~\footnote{Further details of the fitting procedure are given in Ref.~\cite{ColangeloNature2017}}.
Statistics are gathered over 450 repetitions of the experiment, taking into account the inhomogeneous atom-light coupling~\cite{AppelPNAS2009,SchleierSmithPRL2010,ColangeloNature2017}.

The estimate of the state from the two independent measurements is subject to technical noise due to amplitude and phase fluctuations of the input state, and shot-to-shot variations of the magnetic field. In Fig.~\ref{fig:PQS}~a) we plot the estimate of $\F_1$ at time $\tmeas$ for an input state with $\ave{\NA}=1.75\times10^6$ atoms. 
In contrast, the conditional uncertainty of $\F_2$ given $\F_1$ is limited mainly by the measurement read-out noise, as shown in Figs.~\ref{fig:PQS}~b) and c).

From the measurement record we compute the conditional covariance matrix  $\Gamma_{\F_2\mid \F_1}=\Gamma_{\F_2}-\Gamma_{\F_{2} \F_1} \Gamma_{\F_1}^{-1}\Gamma_{\F_1 \F_2}$ which quantifies the error in the best linear prediction of $\F_2$ based on $\F_1$~\cite{BehboodPRL2014}.
$\Gamma_{\bf v}$ indicates the covariance matrix for vector ${\bf v}$, and $\Gamma_{\bf uv}$ indicates the cross-covariance matrix for vectors ${\bf u}$ and ${\bf v}$.
The difference between the best linear prediction of $\F$ using $\F_1$ and the confirming estimate $\F_2$ is visualized using the vector $\bm{\mathcal{F}} = \{ \mathcal{F}_{\rm y}, \mathcal{F}_{\rm z} \} = \tilde{\F}_2 - \Gamma_{\F_{2} \F_1} \Gamma_{\F_1}^{-1} \tilde{\F}_1$, where $\tilde{\F_i}=\F_i-\ave{\F_i}$.
Standard errors in the estimated conditional covariance matrix are calculated from the statistics of $\{\bm{\mathcal{F}}\}$ \cite{Kendall1979}.

Empirically, we find $\tdelta = \unit{270}{\micro\second}$ minimizes the total variance ${\rm Tr}(\Gamma_{\F_2\mid \F_1})$.
This reflects a trade-off of photon shot noise versus scattering-induced decoherence and magnetic-field technical noise.
At this point $\NL = 2.47 \times 10^8$ photons have been used in the measurement and the atomic state coherence has decayed by a factor $\etaSc=0.89$ due to off-resonant scattering, and a factor $\etaDec=0.93$ due to dephasing induced by magnetic field gradients~\cite{ColangeloNJP2013}.
The resulting spin coherence of the PQS is $\FPlane = \etaDec\etaSc\NA$ spins.

\begin{figure}[t]
\includegraphics[width=\columnwidth]{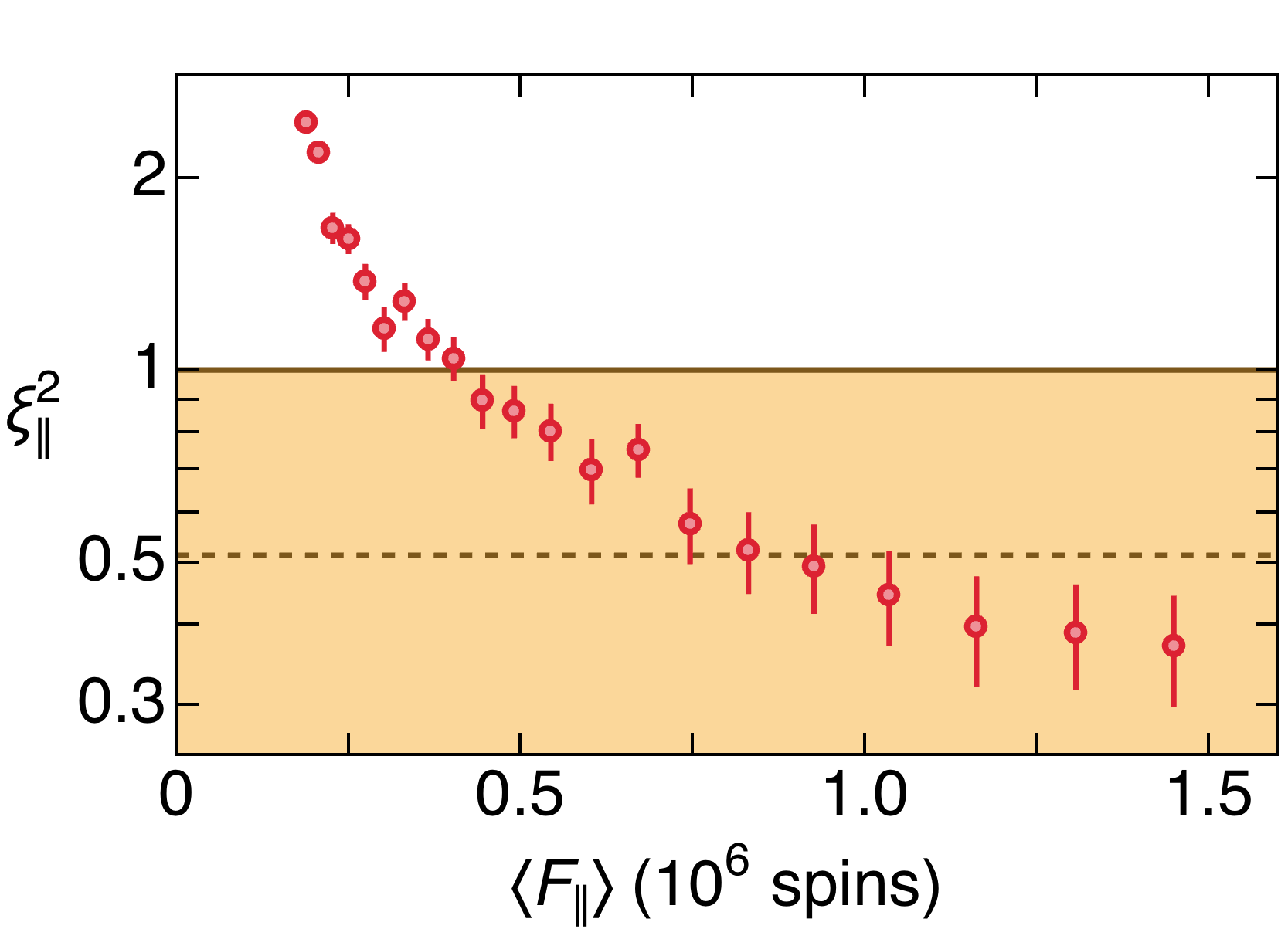}
\caption{
Semi-log plot of the planar squeezing parameter, $\xiPlane$, as function of the in-plane coherence $\FPlane$ of the atomic ensemble.
We vary $\FPlane$ by changing the number of atoms loaded in the optical dipole trap.
A PQS is detected for $\xiPlane<1$ (shaded region).
Entanglement is detected for $\xiPlaneE = (\FPlane/\ave{\tilde{\NA}})\xiPlane <7/16$ (dashed line).
Error bars represent $\pm1\sigma$ statistical errors.
\label{fig:PlotPQS}}
\end{figure}

\begin{figure}[t]
\includegraphics[width=\columnwidth]{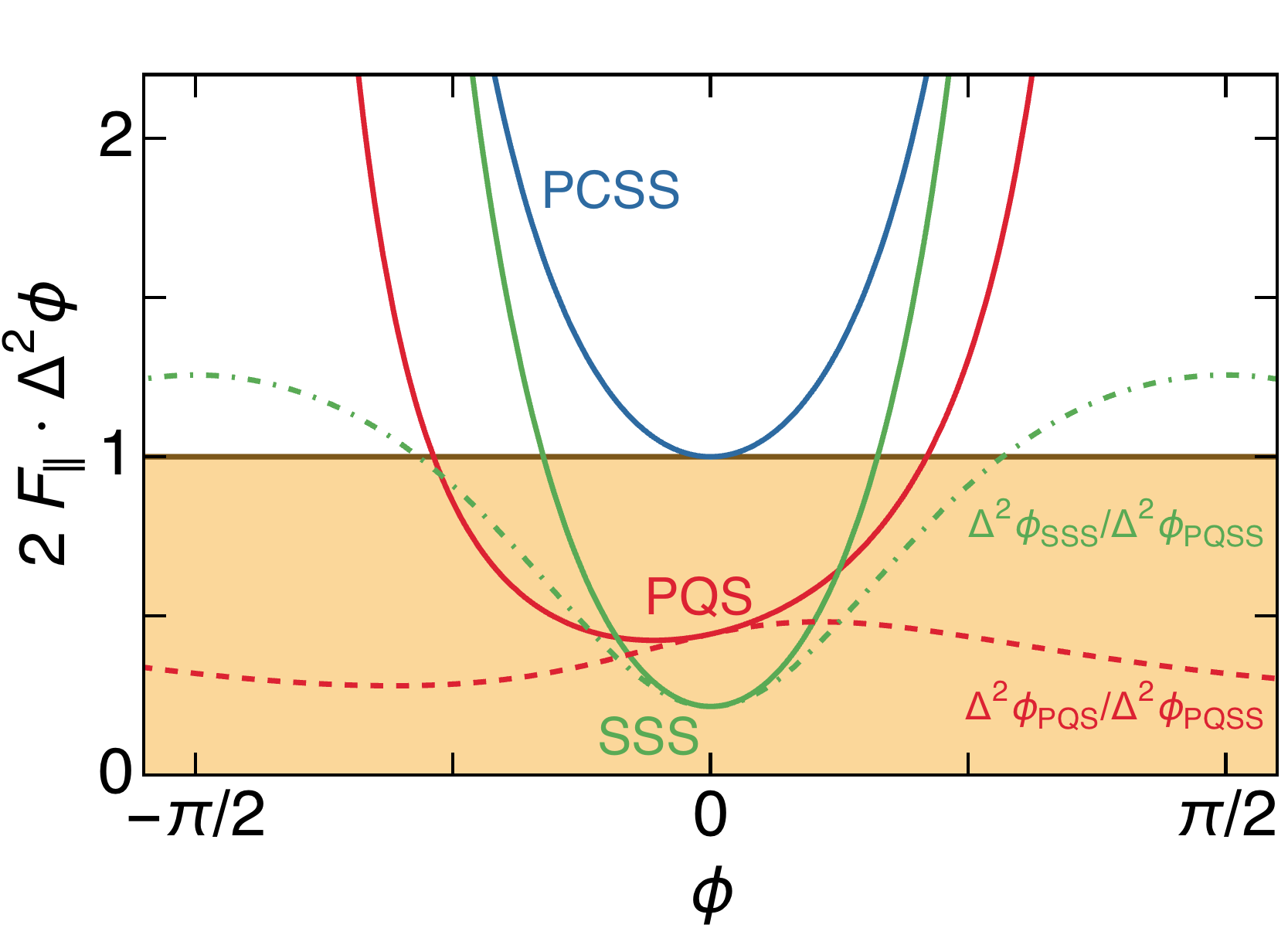}
\caption{
Estimated phase sensitivity of the PQS state as a function of the measurement phase $\phi$ (red solid line).
The standard quantum limit $\Delta^2\phi_{\rm SQL}$ is indicated by the shaded region. 
For comparison, we plot the phase sensitivity of the input PCSS (blue solid line), and an ideal single-variable spin squeezed state (green solid line).
We also show the metrologically significant \textit{enhancement} in phase sensitivity relative to that of the PCSS, $\Delta^2\phi / \Delta^2\phi_{\rm PCSS}$, for both the PQS (red dashed line) and SSS (green dot-dashed line) states.
\label{fig:PLPHASEvsPCSS}
}
\end{figure}

From $\Gamma_{\F_2\mid \F_1}$ we estimate the planar  squeezing parameter $\xiPlane=\Tr(\tilde{\Gamma}_{\F_2\mid \F_1})/\FPlane$, 
where $\tilde{\Gamma}_{\F_2\mid \F_1}=\Gamma_{\F_2\mid \F_1}-\Gamma_{0}$ and $\FPlane$ is estimated at $\tmeas$.
$\Gamma_{0}$ is the read out noise, quantified by repeating the measurement without atoms in the trap.
In Fig.~\ref{fig:PlotPQS} we show $\xiPlane$ as function of the in-plane coherence $\FPlane$ of the atomic ensemble, which we vary by changing the number of atoms in the optical dipole trap.
We detect a PQS for $\FPlane\ge4\times10^{5}$ spins.
With the maximum coherence $\FPlane=1.45\times10^{6}$ spins, we observe $\xiPlane=0.37\pm0.03 < 1$, detecting a PQS with $> 20 \sigma$ significance, with $\xi_y^2 = 0.32\pm0.03$ and $\xi_z^2 = 0.42\pm0.04$, and $\xiPlaneE = 0.32\pm0.02 < 7/16 $, detecting entanglement among the atomic spins with $>5\sigma$ significance~\cite{HePRA2011}.
The measured conditional covariance (in units of spins$^2$) is
\begin{equation}
\Gamma_{\F_2\mid \F_1} =
\left[
\left(
\begin{array}{cc}
 2.32 & 0.64 \\
 0.64 & 3.00 \\
\end{array}
\right)
\pm
\left(
\begin{array}{cc}
 0.21 & 0.16 \\
 0.16 & 0.28 \\
\end{array}
\right)
\right]
\times 10^5.
\label{eq:cov}
\end{equation}
For comparison, the estimated read-out noise is
\begin{equation}
\Gamma_{0} =
\left[
\left(
\begin{array}{cc}
 1.02 & 0.14 \\
 0.14 & 1.03 \\
\end{array}
\right)
\pm
\left(
\begin{array}{cc}
 0.07 & 0.05 \\
 0.05 & 0.07 \\
\end{array}
\right)
\right]
\times 10^5.
\end{equation}

For this state, the observed metrological squeezing parameter is $\xiPlaneM = 0.45\pm0.03$, indicating that entanglement-enhanced phase sensitivity is achievable.
To estimate the enhanced phase sensitivity provided by the PQS state, we evaluate Eq.~(\ref{eq:phase}) using the conditional covariance $\Gamma_{\F_2\mid \F_1}$ and the measured coherences.
The PQS state achieves a maximum sensitivity $\Delta^2\phi = 0.38\, \Delta^2\phi_{\rm SQL}$ ($\Delta\phi=3.6\times10^{-4}$~radians) at a phase $\phi=0.68\,\pi$~radians.
Note that this phase is determined by the choice of measurement time $\tmeas$.

In Fig.~\ref{fig:PLPHASEvsPCSS} we plot the estimated phase sensitivity $\Delta^2\phi$ of the observed PQS state (red solid line).
For comparison purposes, we rotate the PQS so that the spin coherence is aligned along the $y$-axis, 
i.e. $\F \rightarrow R(\theta)\cdot\F$ and $\Gamma_{\F_2\mid \F_1} \rightarrow R(\theta)\cdot \Gamma_{\F_2\mid \F_1} \cdot R(\theta)^{T}$, where $\arctan\theta\equiv\Fy/\Fz$.
We compare this with the sensitivity of a PCSS with input spin coherence $\ave{\Fy}=\NA$ (blue dashed line), and an \textit{ideal} single-variable spin squeezed state (SSS) that would be produced by a single instantaneous quantum non-demolition measurement with the same precision, i.e. with $\Delta^2\Fy = \ave{\NA}$, $\Delta^2\Fz$ reduced by a factor $1/(1+g^2\NL\NA/2)$, and input coherence $\ave{\Fy}=\etaSc\NA$ (green dot-dashed line).

We also plot the calculated \textit{enhancement} in phase sensitivity $\Delta^2\phi$ of both the PQS and SSS states relative to the classical input PCSS.
The measured PQS state achieves $\ge \unit{3.1}{dB}$ quantum-enhanced, metrologically-significant phase sensitivity with respect to the PCSS for all phases, with a maximum of \unit{4.1}{dB}, enabling quantum-enhanced measurement of an \textit{arbitrary} phase shift.
In contract, the SSS achieves 6.6~dB enhancement relative to the PCSS at $\phi=0$, but performs worse than the PQS state outside the range $-0.09\, \pi < \phi < 0.12\, \pi$ radians. 

In contrast to the well known spin-squeezed states, planar quantum squeezed states enhance the precision of phase estimation without requiring \textit{a priori} information about the phase.  Here we have shown that QND measurement can efficiently produce such states, demonstrating more than \unit{3}{dB} of advantage relative to classical states over the full range of phase angles. We also detect spin-spin entanglement underlying the metrological advantage.  Such states are  attractive for high-bandwidth and high-sensitivity optical magnetometers~\cite{KominisN2003,ShahPRL2010} and other atomic sensing applications employing non-destructive spin detection \cite{LodewyckPRA2009,ShengPRL2013,HostenN2016}.

\begin{acknowledgments}
\section*{Acknowledgements}
We thank Q.~Y. He, M. Reid, P. Drummond, G. Vitagliano, G. T\'oth, E. Distante, V.G. Lucivero, L. Bianchet, N. Behbood and M. Napolitano for helpful discussions.
Work supported by MINECO/FEDER, MINECO projects MAQRO (Ref. FIS2015-68039-P), XPLICA (FIS2014-62181-EXP) and Severo Ochoa grant SEV-2015-0522, Catalan 2014-SGR-1295, by the European Union Project QUIC (grant agreement 641122), European Research Council project AQUMET (grant agreement 280169) and ERIDIAN (grant agreement 713682), and by Fundaci\'{o} Privada CELLEX.
GP gratefully acknowledges funding from the Agencia Nacional de Promoci\'{o}n Cient\'{i}fica y Tecnol\'{o}gica (ANPCyT), PICT2014-1543, PICT2015-0710, and UBACYT PDE 2017.
\end{acknowledgments}


%

\end{document}